\documentclass[pra,reprint,longbibliography,superscriptaddress]{revtex4-1}
\usepackage{amsmath}
\usepackage{graphicx}
\usepackage{epstopdf}
\usepackage{color}

\begin{document}

\title{Unusual quasiparticles and tunneling conductance in quantum point contacts in $\nu=2/3$ fractional quantum Hall systems}
\author{Vadim Ponomarenko}
\affiliation{Ioffe Institute, 194021, St. Petersburg, Russia}
\author{Yuli Lyanda-Geller}
%\email{yuli@purdue.edu}
\affiliation{Department of Physics and Astronomy, Purdue University, West Lafayette, IN 47907 USA}
\affiliation{Quantum Science and Engineering Institute, Purdue University, West Lafayette, IN 47907 USA}
%\email{yuli@purdue.edu}
%\date{\today\ @ \currenttime, ---= draft: \jobname.tex =---}
\date{\today}

\begin{abstract}
Understanding topological matter in the fractional quantum Hall  (FQH) effect requires identifying the nature of edge state quasiparticles.
FQH edge state at the filling factor $\nu=2/3$ in the spin-polarized and non-polarized
phases is represented by the two
modes of composite fermions (CF) with the parallel or opposite spins described by the chiral Luttinger liquids.
Tunneling through a quantum point contact (QPC) in such systems between different or similar spin phases is solved exactly. With the increase of the applied voltage, the QPC conductance grows from zero and saturates at $e^2/2h$ while a weak electron tunneling between the edge modes with the same spin transforms into a backscattering carried by the charge $q=e/2$ quasiparticles. These unusual quasiparticles and conductance plateau emerge in the QPC with one or two CF modes scattering into a single mode, as occurs in these systems.
We propose experiments on the applied voltage and temperature dependence of the QPC conductance and noise that can shed light on the nature of edge states and FQH transport.
\end{abstract}

\maketitle
\vspace{-2mm}
Quantum Hall effect (QHE) emerges in quantizing magnetic fields in a two-dimensional electron gas. It exhibits a zero longitudinal resistance and the Hall conductance quantized in terms of the number of conductance quanta $e^2/2\pi\hbar$, where $e$  is the electron charge and $\hbar$ is the Planck constant.
When electron correlations become important, the fractional quantum Hall effect (FQHE) exhibits a fractionally quantized conductance. FQHE excitations are fractionally charged quasiparticles that obey abelian or non-abelian anyon statistics \cite{lm77,wilczek1982a,wilczek1982,wilczek83,Halperin1984,Arovas1984,Read2000}.
FQHE with its edge states is the remarkable breeding ground for the observation of topological phenomena and the creation of qubits for topological quantum computing \cite{Nayak2008, Lindner2012,Clarke2012,Vaezi2013,Mong2014,Vaezi2014,Simion2018, Liang2019,Scarola2019,Tylan2015}.

QHE conductance quantization in many cases stems from the edge state picture \cite{halperin82} in the Landauer- Buttiker approach \cite{buttiker88}. For the FQHE, Wen \cite{Wen1995} found that at filling factors of Laughlin \cite{laughlin83} states $\nu=n\Phi_0/B=1/m$, where $m$ is odd, $n$ is the electron density, $\mathbf{B} $ is the external magnetic field, and $\Phi_0=2\pi\hbar c/e$ is the flux quanta, the excitations on the edge of the 2D electron system can be described as the chiral Luttinger liquids.
While in quantum wires \cite{Ponomarenko1995,Maslov1995,Safi1995} conductance does not depend on the Luttinger liquid parameters, the picture of chiral Luttinger liquids allowed to explain the quantized Hall conductance \cite{Wen1995,Kane1992} and tunneling of $\nu=1/m$ edge states through the quantum point contact (QPC)  \cite{Fendley1995,Fendley1995a,Sandler1998}. Furthermore, in \cite{Fendley1995} the thermodynamic Bethe Anzatz and the integrability
of the Luttinger model with impurity interaction were used  to calculate the non-equilibrium temperature-dependent conductance and noise.

A more complex situation emerges for hierarchical FQHE states \cite{Haldane1983,Girvin1983,Halperin1984}, such as
$\nu=2/3$ state, for which
 several models were proposed. Since this state
is a hole-conjugate of the $\nu=1/3$ state, it was suggested in \cite{Macdonald1990, Johnson1991} that
 its edge state consists
of an outer downstream integer mode (from the underlying $\nu=1$ state) and an inner, counterpropagating -1/3 edge mode. Consideration of a bilayer system in \cite{MacHaldane} discussed this conjugated state  and two other models,  two independent $\nu=1/3$ edge channels and a pseudospin singlet state.  The importance of disorder scattering between edge modes was underscored in \cite{Kane1994a,Kane}, where it was shown that the scattering
results in two decoupled modes, a single downstream charge
mode with conductance $2e^2/3h$ and a counterpropagating neutral mode. Soft confining potential  leads to a reconstructed edge involving the formation of $\nu=1/3$ strip and two downstream charge modes with
conductance $e^2/3h$ and counterpropagating neutral
modes that were considered in \cite{Meir1,Meir2}.
With different approaches to the structure of the edge, there have also been two approaches to transport. Conductance quantization implies equilibration between the chemical
potentials of the reservoirs and the outgoing edges. The first approach, incoherent transport  model, assumes a short coherent length and suppression of quantum interference between
the channels \cite{KF,SF,MG,Nosiglia2018,Spanslatt}; the second approach is quantum solution taking interference into account \cite{Fendley1995,PA}.
Experimentally, studies of tunneling through quantum point contacts (QPC) at $\nu=2/3$ in \cite{Bid2009,Baer,Sabo} have shown a conductance  plateau at $e^2/3h$, however, recent work \cite{Manfra} demonstrated its appearance within $2\%$ of  $e^2/2h$.

In this paper, we find an exact quantum mechanical solution to the model of tunneling through QPC  in several $\nu=2/3$ edge configurations. It exhibits the fractional charge $e/2$ of backscattering quasiparticles, and $ e^2/2h$ conductance  plateau. One case when the model is realized is associated with the transition between spin-polarized and unpolarized $\nu=2/3$ FQH phases \cite{Eisenstein1990,Kraus2002,Wu2012a} which occurs due to the crossing of the two composite fermion (CF) energy states \cite{Jain1989,JainCFbook2007} with opposite spin polarization \cite{Ronen2018,Kukushkin1999,Liang2019}. It was recently shown  \cite{Kazakov2016,Kazakov2017,Wu2018,Wang2021} that  the two different spin phases can be induced in the neighboring regions of the FQH liquid by electrostatic gates. The properties of a contact between the regions naturally lead to the model with two CF edge modes of the spin-polarized phase coupled by electron tunneling through QPC at low voltage and temperature to one CF edge mode of the same spin in the unpolarized phase.
The same model also describes tunneling through QPC between the FQH regions in the identical spin phases when two CF edge modes in one of those regions are well separated in space due to its potential profile so that only one mode can tunnel. The exact solution to the model shows that with an increase of the applied voltage or temperature, the QPC conductance grows from zero and saturates at $e^2/2h$ while a weak electron tunneling between the edge modes with the same spin transforms into the backscattering carried by the charge $q=e/2$ quasiparticles. This confirms the $e^2/2h$ conductance earlier found by us for the contact between spin-polarized and unpolarized regions using a strong coupling boundary conditions approach \cite{Wang2021}. We show that the quasiparticle $e/2$ charge is also observable in the shot noise.
Testing the predicted non-linear applied voltage dependence of the tunneling current, temperature dependence, and properties of the shot noise can distinguish coherent and incoherent transport models and uncover the physics of edges in both polarized and unpolarized FQHE at $\nu=2/3$.

{\it Model for edge states.} The key point of our approach is the assumption of separation of the charge and neutral/spin  edge modes due to the long-range Coulomb interaction, discussed in \cite{Wen1995}. We treat polarized and unpolarized phases of $\nu=2/3$ FQHE on equal footing describing edge states as chiral Luttinger
liquids with action 
\begin{eqnarray}
S = \frac{1}{4\pi}\int dt \int dx \left[ -3\partial_x \varphi_{c}(\partial_t+v_c\partial_x) \varphi_{c}\right.   +\nonumber\\
  \left.\partial_x \varphi_{n}   (\partial_t- v_n\partial_x) \varphi_{n}\right],
\label{action}
\end{eqnarray}
where $\varphi_c$ and  $\varphi_n$ are bosonic operators and  $v_c$ and $v_n$ are velocities  of the charge and the neutral modes, correspondingly. Their relation to electron edge modes corresponding to the two CF $\Lambda-$ levels and to quasiparticle bosonic fields, properties of these operators, transformations between different representations, and underlying assumptions are discussed in the supplementary materials. The neutral operators describe the difference in the occupation numbers of
$\Lambda$-levels in the polarized phase and the spin density in the unpolarized phase. We will use the notation $\varphi_{pc}$ and $\varphi_{uc}$ for charge modes, $\varphi_{pn}$ and $\varphi_{us}$ for neutral (spin) modes in the corresponding phases. We assume that no edge reconstruction takes place.

{\it Tunneling and charge current.}
We begin with tunneling through a QPC  between the polarized and unpolarized  phases.
In the weak coupling limit, it is carried by
the whole electrons through the allowed spin-conserving processes involving the edge modes of the two CF $\Lambda$ levels in each phase. The point contact (junction) electron tunneling at $x=0$ between polarized and unpolarized phases is described by tunneling Hamiltonian
\begin{equation}
{\cal H}_T=-\sum_{a=1,2}\large( U_a \xi_u \xi_{pa}e^{i(\Psi_a(t)-Vt)} + h.c.\large).
\label{tunHam}
\end{equation}
Here $\{\xi_b,\xi_c\}=2\delta_{b,c}$, $\xi$'s are Majorana fermions accounting for the fermion statistics of
different CF edge modes and composing the corresponding Klein factors, and
\begin{eqnarray}
&\Psi_a(t)=\Phi_{pa}(0,t)- \Phi_{u1}(0,t)\nonumber \\
&=\frac{1}{\sqrt{2}}(3\varphi_{pc}(0,t)\mp \varphi_{pn}(0,t)-3\varphi_{uc}(0,t)+ \varphi_{us}(0,t)),
\label{tildePhi}
\end{eqnarray}
where $\Phi_{pa}$ and $ \Phi_{u1}$ are the CF fields, the tunneling amplitudes $U_a$ can be chosen real and positive, and $V$ is the applied voltage. We assume coherent propagation of charge and neutral (spin) modes along the edge and allow for the possibility of interference.
The same tunneling Hamiltonian Eq.~(\ref{tunHam}) can also describe tunneling between the two spin-polarized phases, with a single edge channel $u=p^{\prime}$ participating in tunneling, if the potential profile ensures that the second $\Lambda$-level edge
is located too far from the $p^{\prime}$ edge. Thus we study the QPC tunneling of two modes  into one mode; tunneling of one mode into one mode will emerge for tunneling amplitude $U_2=0$.

The tunneling charge current, from  Eq.~(\ref{tunHam}) is given by
\begin{equation}
J_T=-\partial_t \hat{Q_P}(t)=i\left[\int dx \rho_{pc}, {\cal H}_T\right]= \frac{\delta}{\delta(Vt)} {\cal H}_T,
\label{tuncurrent}
\end{equation}
where $\rho_{pc}=\frac{1}{\sqrt{2}\pi}\partial_x\varphi_{pc}$ is the charge density, e=1.
Its average and fluctuations are defined by the operators $\Psi_{a}(t), \ a=1,2$, whose evolution
in the absence of tunneling is characterized by their second-order correlators. The latter are
found from the Gaussian action Eq.(\ref{action}) for $\varphi_{pc}$, $\varphi_{pn}$ and similarly for $\varphi_{uc}$ and $\varphi_{ps}$ in the form
\begin{equation}
<\!\Psi_a(t)\Psi_b(0)\!>=(3+\delta_{a,b})g(t), \label{g}
\end{equation}
$g(t)$ is the single point correlator at $x=0$ of a normalized chiral boson field, with action as in Eq.(\ref{action}) for
$\varphi_{pn}$,
\begin{equation}
<\!\varphi_{pn}(x,t)\varphi_{pn}(0,\!0)\!\!>=\!-\ln(\delta (i(t\!+\!\frac{x}{v_n})\!+\!\alpha))\!\equiv g(t\!+\!\frac{x}{v_n}).
\end{equation}
Here $\alpha = 1/D$, $D$ is the energy cut-off in both edges, and $\delta\rightarrow 0$ should be taken in the final results. In terms of the  normalized chiral boson fields $\phi_{j}(x,t), j=0,a$,
the operators $\Psi_{a}(t), \ a=1,2$ can also be represented as
\begin{equation}
\Psi_a(t)=\phi_a(0,t)+\sqrt{3}\phi_0(0,t) ,
\end{equation}
which after fermionization
\begin{equation}
\psi_a(x,t)\!=\!\frac{\xi_{pa}}{\sqrt{2\pi \alpha}} e^{i\phi_a(x,t)}\!, \  \psi_0(x,t)\!=\!\frac{\xi_{u}}{\sqrt{2\pi \alpha}} e^{i\sqrt{3}\phi_0(x,t)}
\end{equation}
allows to re-write the tunneling Hamiltonian (\ref{tunHam}) as
\begin{equation}
{\cal H}_T\!=\!-\!\!\sum_{a=1,2}\large( 2\pi \alpha U_a \psi^+_0(0,t) \psi_a(0,t)e^{-iVt}\!\!+\!h.c.\large),
\end{equation}
leading to its interpretation as the tunneling process of electrons from the two non-interacting
chiral channels into the FQHE edge of $\nu=1/3$ filling factor. Combining these interfering channels into a single
tunneling channel and applying bosonization of its chiral fermion field
\begin{equation}
\psi_{T}(x,t)=\frac{1}{\sqrt{\sum_a U_a^2}} \sum_a U_a \psi_a(x,t)\equiv \frac{\xi_{T}}{\sqrt{2\pi \alpha}} e^{i\phi_T(x,t)}
\end{equation}
we obtain the tunneling Hamiltonian ${\cal H}_T$ and current $J_T$:
\begin{eqnarray}
{\cal H}_T= -\tilde{t}\cos{(\phi_T(0,t)-\sqrt{3}\phi_0(0,t)-Vt)}
\label{tunnelingH} \\
  J_T=-\tilde{t}\sin{(\phi_T(0,t)-\sqrt{3}\phi_0(0,t)-Vt)},
\label{tunnelingc}
\end{eqnarray}
where $\tilde{t}^2=2\sum_a U_a^2$, and the single Klein factor is omitted since its drops out from any perturbative order in $\tilde{t}$ due to the charge conservation.

The electron tunneling described by  Eqs.~(\ref{tunnelingH},\ref{tunnelingc}) can be viewed as a single point electron tunneling
between the two counterpropagating  primary one-component edges of equal filling factor $\nu=1/2$ with densities
$\rho_{d,n} (x,t) =\pm \frac{1}{2\pi}\sqrt{\nu}\partial_x\phi_{d,n}(x,t)$ and the corresponding electron annihilation operators $e^{i\frac{1}{\sqrt{\nu}}\phi_{d,n}(x,t)}$.
Here $\phi_{d,n}$ are normalized right and left moving chiral fields.  Thus, the problem described by Eq.(\ref{tunHam}) maps onto and is equivalent
to the problem of tunneling between the one-component edges $\phi_{d}(x,t)$ and $\phi_{n}(x,t)$ as described by  Hamiltonian
\begin{equation}
{\cal H}_T=-\tilde{t}\cos{\left(\frac{1}{\sqrt{\nu}}\phi_{d}(0,t)- \frac{1}{\sqrt{\nu}}\phi_{n}(0,t)-Vt\right)}.
\label{SX}
\end{equation}

\begin{figure}
\includegraphics[width=0.49\textwidth]{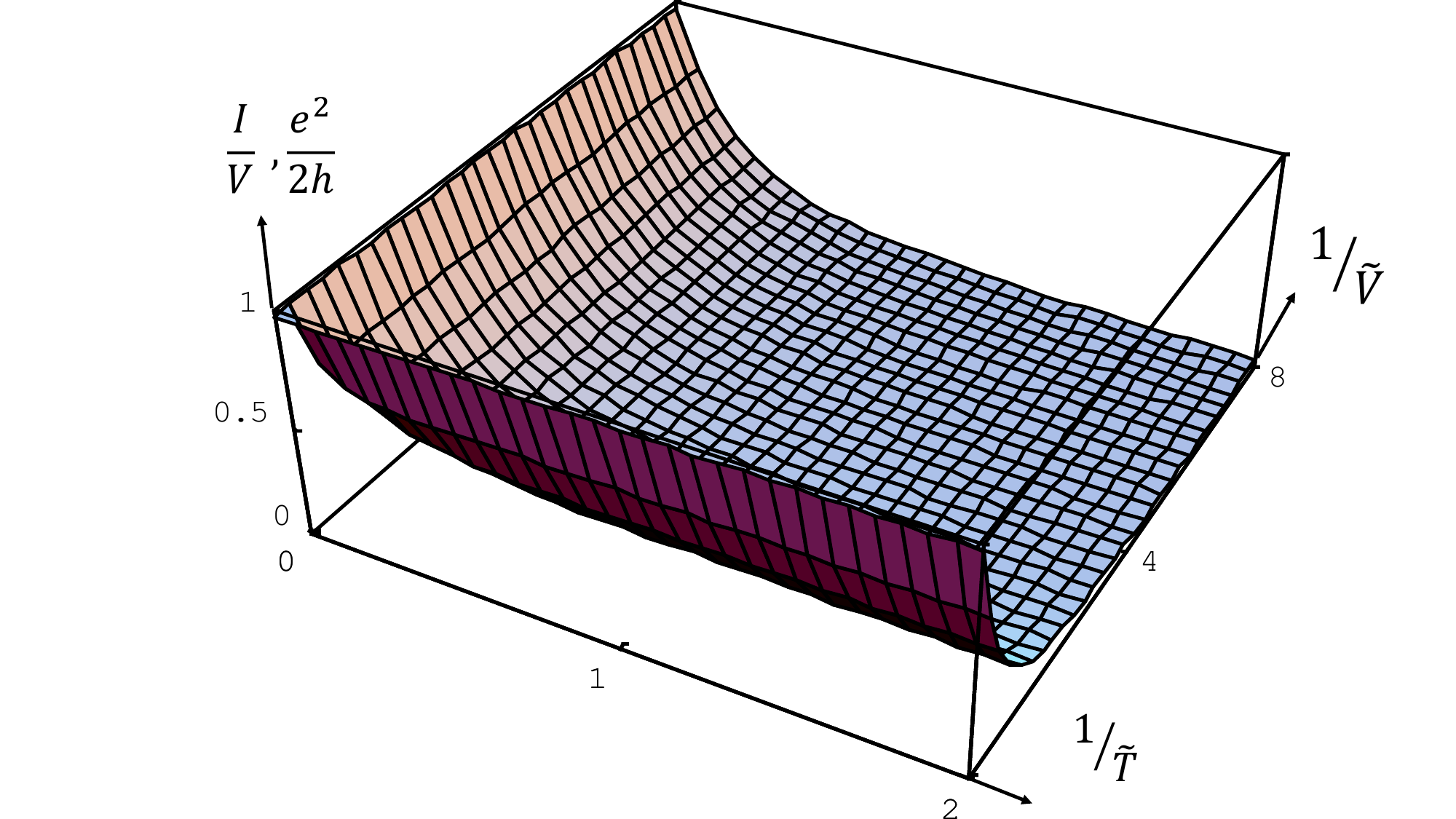}
\vspace{-0.4cm}
\caption{
\label{3D}
Conductance $\langle J_T \rangle/V=I/V$ dependence  on dimensionless variables $1/\tilde{V}=\Gamma/V$ and $1/\tilde{T}= \Gamma/2 \pi T.$} 
\end{figure}

The tunneling current $J_{T}$ in the latter problem coincides with $J_T$  in Eqs.~(\ref{tuncurrent},\ref{tunnelingc}).  Its average was calculated in \cite{Fendley1995}. At finite temperatures $T$ this average  is given by
\begin{equation}
\langle J_T \rangle= \frac{1}{4\pi}\left(V-\frac{\Gamma^2}{2}\int d\omega \frac{f(\frac{\omega-V}{T})- f(\frac{\omega+V}{T})}{\omega^2 + \Gamma^2}\right),
\label{current_general}
\end{equation}
where $f(\frac{\epsilon}{T})$ is the Fermi distribution function, and $\Gamma=2D^2/(\pi \tilde{t})$ \cite{Weiss}.
The integral in (\ref{current_general}) is evaluated in terms of the imaginary part of the digamma-function $\psi(x)$ \cite{Weiss1991}:
\begin{equation}
\langle J_T \rangle=\frac{1}{4\pi}\left(V-\Gamma \  {\rm Im}\psi\!\left[ \frac{1}{2}+ \frac{\Gamma+iV}{2\pi T}\right]\right),
\label{eval}
\end{equation}
describing the dependence of tunneling current on the applied voltage and temperature in the whole range of parameters. In  Fig.\ref{3D} conductance $\langle J_T \rangle/V$  is plotted as function of dimensionless variables $1/\tilde{V}=\Gamma/V$ and $1/\tilde{T}= \Gamma/2\pi T$. At $T=0$ $\langle J_T \rangle$ is given by
\begin{equation}
\langle J_T \rangle= \frac{1}{4\pi}\left( V- \Gamma \  \tan^{-1}\!{\left(\frac{V}{\Gamma}\right)}\right).
\label{current0}
\end{equation}

{\it Tunneling current in various regimes.}
In order to elucidate different contributions to the tunneling current, we consider limiting cases of Eq. (\ref{current0}).
In the regime of low $V$ , $V\ll \Gamma$, of electron tunneling described by Eq. (\ref{SX}), the current $\langle J_T^{l} \rangle$ is cubic in $V$:
\begin{equation}
 \langle J_T^{l} \rangle= \frac{1}{3}\frac{V}{4\pi}\left(\frac{V}{\Gamma}\right)^2.
\end{equation}
Therefore, we observe that in the coherent approach to transport, the behavior of tunneling current and conductance at small voltages is nonlinear in the applied voltage for $T\ll V$, as opposed to the picture of incoherent transport \cite{MG,Nosiglia2018,Spanslatt} that results in Landauer-Buettiker current linear in voltage and voltage-independent conductance.

In the regime of high voltages,  $V\gg \Gamma$, the exact Eq.(\ref{current0})  gives the following result for the average current, which we will denote $J_T^{h}$:
  \begin{equation}
\langle J_{T}^{h}\rangle=\frac{V}{4\pi} - \frac{\Gamma}{8}.
\end{equation}
The first term in this expression defines conductance $G=e^2/2h$. The second term corresponds to the reduction of the tunneling current $\langle J_T\rangle$ defined by by the quasiparticle backscattering current $J_{bsc}$
\begin{equation}
J_{bsc}=\langle J_T\rangle- \langle J_{T}^{h}\rangle=\frac{\Gamma}{8},
\end{equation}
where $\langle J_T\rangle=V/4\pi$.

In the supplement, we show an alternative way to calculate the tunneling currents using the strong coupling boundary conditions. We calculate the quasiparticle charge, and neutral and spin tunneling currents
for tunneling through QPCs between polarized and unpolarized phases or between two polarized FQH liquids with one edge channel precluded from tunneling. The fermionization method above shows that tunneling of two interfering edge channels from one QPC side into a single channel on the other side is equivalent to a single tunneling channel with a renormalized tunneling amplitude.

We note that QPC conductance $G=e^2/2h$ has been also discussed \cite{Manfra} in terms of the incoherent equilibration model. However, neither the nonlinear current-voltage characteristics nor the temperature dependence following from 
Eq. (\ref{eval}) arise in that approach. $G=e^2/2h$ has been also discussed for a special tunneling configuration, in which tunneling between two \emph{$\nu=2/3$} regions proceeds through the quantum dot with $\nu=1$ \cite{KYang}.

{\it Quasiparticle charge}. The reduction of tunneling current due to quasiparticle backscattering is described and their charge  is determined by taking into account a sudden change of strong coupling boundary conditions at $t=t_0$. For tunneling, e.g., between $p1$ edge mode in the polarized region and $u1$ edge mode with the same spin in the   unpolarized region through the QPC  at $x=0$ the strong coupling boundary conditions  $\tilde{t}\rightarrow \infty$ are given by
\begin{equation}
\frac{1}{2}\sum_{\pm}\left(  \tilde{\Phi}_{p1}(\pm0,t_0) -\tilde{\Phi}_{u1}(\pm0,t_0)\right)=2\pi n.
\end{equation}
Here both $\Phi_{\alpha ,1}(x,t), \alpha=p,u$ fields are extended to finite $x$ from their $x=0$ expressions in Eq. (\ref{tildePhi}), as chiral right moving fields.
A sudden variation of the boundary conditions changes $n=0$ to $n=1$; both $n$ minimize $-\tilde{t}\cos{(\Phi_{p1}(0,t)- \Phi_{u1}(0,t))}$. The boundary conditions also must keep continuous the two dual fields
\begin{eqnarray}
 &\eta_{1}(x,t) =\Phi_{p1}(x,t)\theta(-x)+\Phi_{u1}(x,t)\theta(x)\\
&\eta_{2}(x,t) =\Phi_{u1}(-x,t)\theta(-x)+\Phi_{p1}(x,t)\theta(x),
\label{dual}
\end{eqnarray}
where $\theta(x)$ is the Heaviside step function.
The jump from $n=0$ to $n=1$ leads to a jump in the dual fields
\begin{equation}
\left(\eta_1(x,t_0)- \eta_2(x,t_0)\right)\vert_{x=-0}^{+0}=-4\pi.
\end{equation}
Since $\left(\eta_1(x,t_0)+\eta_2(x,t_0)\right)\vert_{x=-0}^{+0}=0,$
%\begin{equation}
%\left(\eta_1(x,t_0)+\eta_2(x,t_0)\right)\vert_{x=-0}^{+0}=0,
%\end{equation}
the corresponding jumps in these fields are given by
\begin{equation}
\Delta\eta_2(+0,t_0)=2\pi= -\Delta\eta_1(+0,t_0).
\end{equation}
Then we get the change
\begin{equation}
\Delta\left(3\varphi_{pc}(x,t_0)-\varphi_{pn}(-x,t_0)\right)= 2\sqrt{2}\pi\theta(x).
\end{equation}
Using the continuity condition
\begin{equation}
\Delta\left(\varphi_{pc}(x,t_0)+\varphi_{pn}(-x,t_0)\right)=0,
\end{equation}
we find
\begin{equation}
\Delta\varphi_{pc}(x,t_0)=\frac{\pi}{\sqrt{2}}\theta(x),\hspace{4mm}
\Delta\varphi_{pn}(-x,t_0)=-\frac{\pi}{\sqrt{2}}\theta(x).
\end{equation}
Then the changes in the charge density and  the density of the neutral mode are giiven by 
\begin{equation}
\label{charges}
\delta\rho_{pc}=\frac{1}{\sqrt{2}\pi}\partial \Delta\varphi_{pc}=\frac{1}{2}\delta(x),\hspace{3mm}
\delta\rho_{pn}=\frac{1}{2}\delta(x).
\end{equation}
In the supplement, we evaluate changes in tunneling densities of the spin mode in the unpolarized region and demonstrate relations between changes of densities for charge, neural and spin modes on both sides of the QPC.

As was shown in \cite{Ponomarenko2004}, the calculated transferred charge (\ref{charges}) enables us to identify the charge $q$ of backscattering quasiparticles. Here $q=e/2$. Surprisingly, this charge is different from $q=e/3$ that one expects from \cite{Kane1994a}. This difference stems from the absence of tunneling between two edge states of opposite spin polarizations in the polarized and unpolarized regions, or, for tunneling between the regions with the same spin polarization, from a single channel participating in tunneling in at least one of the regions, due to a potential profile.

{\it Shot noise.}
The experimental determination of the charge of the tunneling quasiparticles can be achieved by measurements of the shot nonequilibrium noise \cite{Noise1,Noise2}. The shot noise of the tunneling current \cite{Kane1994,Fendley1995,Chamon1996,Lesage1997,Sandler1999} is given by
 \begin{equation}
S=\int_{-\infty}^{+\infty}{ dt \left( \langle J_T(t)J_T(0)\rangle-\langle J_T\rangle^2\right)} .
\end{equation}
From \cite{Fendley1995} it follows that
\begin{eqnarray}
&S=-\frac{\nu}{2(1-\nu )}V^2 \partial_V\left(\frac{\langle J_T\rangle}{V}\right)=\nonumber \\
&\frac{1}{2}\frac{\Gamma}{4\pi}\left( \tan^{-1}{\left(\frac{V}{\Gamma}\right)}-\frac{v/\Gamma}{1+ \left(\frac{V}{\Gamma}\right)^2}\right).
\end{eqnarray}
At small voltages and weak backscattering $V\ll\Gamma$
\begin{equation}
S^{l}= \frac{1}{3}\frac{V}{4\pi}\left(\frac{V}{\Gamma}\right)^2.
\end{equation}
At large voltages $V\gg\Gamma$
\begin{equation}
S^{h}=\frac{1}{2}\frac{\Gamma}{4\pi}\frac{\pi}{2}.
\end{equation}
The Schottky formula gives the charge of tunneling quasiparticles $q$  in the limits of weak electron tunneling and weak quasiparticle backscattering as the ratio of the corresponding values of noise to the currents, $S^{l}/\langle J_T \rangle$   and $S^{h}/\langle J_{bsc} \rangle$.
The resulting charge (writing e explicitly) is
\begin{equation}
S^{l}/\langle J_T \rangle= q=e,\hspace{3mm}
S^{h}/\langle J_{bsc} \rangle=q=e/2.
\end{equation}
Thus, the result for shot noise confirms the one obtained from the analysis of the sudden change in the boundary conditions: for tunneling at a filling factor $\nu=2/3$ through the QPC in our model, the quasiparticle charge is
$e/2$ rather than the expected $e/3$. We note that the noise signal arising in our model is much stronger than the noise in the incoherent equilibration model.

{\it Discussion and conclusion.}
We have shown that the coherent quantum-mechanical model of transport in the $\nu=2/3$ FQHE, that potentially involves interference of chiral Luttinger liquid edge modes, leads to an exact solution of the problem of tunneling through the QPC between $\nu=2/3$ FQHE regions with different or similar spin phases.
With the increase of the applied voltage, the QPC conductance grows from zero and saturates  at $e^2/2h$ while a weak electron tunneling through the QPC between the same spin edge modes transforms into the backscattering carried by the fractional charge $q=e/2$ quasiparticles. Unusual new quasiparticles and fractional conductance emerge in the QPC with one or two CF modes scattering into one mode as is the case in tunneling between polarized and unpolarized FQHE phases and can occur in tunneling between similar phases as a result of engineering the potential profile. Using the fermionization method, we have shown that
 tunneling of the two CF modes from one of the sides of the QPC is renormalized with the account of interference between the two modes, and is equivalent to tunneling of a single mode.

Recent experiments on QPC tunneling at $\nu=2/3$ have shown signatures of $e^2/2h$ plateau. Besides our model  this conductance can be discussed using the incoherent equilibration transport approach. However, then the non-linear current-voltage characteristics and temperature dependence of tunneling current and noise, which are predicted in the present work and can be tested experimentally, do not emerge in the incoherent model.

{\it Acknowledgement.}
We are grateful to M. Manfra and L. Rokhinson for helpful discussions.  YLG was supported by the U.S. Department of Energy, Office of Basic Energy Sciences, Division of Materials Sciences and Engineering under Award DE-SC0010544.

\bibliography{tunneling12}

%--------------------------------------------------------------------------------------------
% Supplement
%--------------------------------------------------------------------------------------------

\renewcommand{\thefigure}{S\arabic{figure}}
\renewcommand{\theequation}{S\arabic{equation}}
\renewcommand{\thepage}{sup-\arabic{page}}
\setcounter{page}{0}
\setcounter{equation}{0}
\setcounter{figure}{0}

\begin{center}
\textbf{Supplementary Materials}
\end{center}

\section{Note on scope of the supplemental materials.}
 
In the supplemental materials we discuss relations between different representations of edge tunneling modes, paricularly 
between electron edge modes corresponding to the two CF $\Lambda-$ levels and quasiparticle bosonic fields, transformations between different representations, and underlying assumptions of our model. We also show an alternative way to calculate the tunneling currents using the strong coupling boundary conditions and calculate the quasiparticle charge, and neutral and spin tunneling currents
for tunneling through QPCs between polarized and unpolarized phases or between two polarized FQH liquids with one edge channel precluded from tunneling.
Finally, in addition to calculation of tunneling density of the charge mode in the main text, we  evaluate changes in tunneling densities of the spin mode in the unpolarized region and demonstrate relations between changes of densities for charge, neural and spin modes on both sides of the QPC.

\vspace{5mm}
{\it Notations used in the supplement.}\\

 All equations numbers of the supplementary materials are labeled by capital S. Citations numbers refer to the list of references in the main text.  

\section{ Modeling of tunneling through a quantum point contact at $\nu=2/3$}

\subsection{Description of edges of $\nu=2/3$ state in terms of bosonic fields and quasiparticle bosonic fields}

Luttinger liquid action for $ \nu=2/3$ edge states in terms of bosonic fields of electron operators $\Phi_{1,2}$ is written as
\begin{equation}
S = -\frac{1}{4\pi}\int dt \int dx \left[ \partial_x \Phi \hat{K}^{-1}\partial_t\Phi   +  \partial_x \Phi\hat{V}^{e}  \partial_x \Phi\right],
\end{equation}
and the charge density is
\begin{equation}
\rho_c(x)= \frac{1}{2\pi}\left(q\hat{K}^{-1}\partial_x\Phi (x)\right)=\frac{1}{2\pi}\frac{1}{3} \left(\partial_x\Phi_1+\partial_x\Phi_2\right),
\end{equation}
where matrix $K$ and vector q are given by
\begin{equation}
\mathbf{K}=\left(\begin{array}{cc}1 & 2\\2 & 1 \end{array}\right),
\mathbf{q}=\left(\begin{array}{c}1 \\1 \end{array}\right).
\label{K}
\end{equation}

In the most general form the interaction matrix $\hat{V}^{e}$ is defined by two different diagonal matrix elements $V_1$ and $V_2$for intra-mode Coulomb interactions in the modes $\Phi_{1,2}$  and the
off-diagonal matrix element $V_3$ for inter-mode Coulomb interactions,

\begin{equation}
\hat{V}^{e}=\left(\begin{array}{cc}V_1 & V_3\\V_3 & V_2 \end{array}\right).
\label{V}
\end{equation}

The commutation relation is
\begin{equation}
\left[ \partial_x \Phi_i(x), \Phi_j(x')\right]=i2\pi K_{ij}\delta(x-x').
\end{equation}
For composite fermion creation operators, we have
\begin{equation}
\Psi^{\dagger}_j (x)\propto \exp{(-i\Phi_j(x))}
\end{equation}
and
\begin{equation}
\left[\rho_c(x), \Psi^{\dagger}_j(x')\right]=\delta(x-x')\Psi^{\dagger}_j(x').
\end{equation}

Luttinger liquid action in terms of quasiparticle bosonic fields $\chi$ , defined by $\Phi = {\hat K}\chi$ is given by
\begin{equation}
S = -\frac{1}{4\pi}\int dt \int dx \left[ (\partial_x \chi \hat{K}\partial_t\chi )  +
 (\partial_x \chi\hat{V}^{qp}  \partial_x \chi)\right],
\end{equation}
where
\begin{equation}
\hat{V}^{qp}=\hat{K}\hat{V}^{e}\hat{K}.
\end{equation}
The two sets of fields are orthogonal:
\begin{equation}
\left[ \partial_x \chi_i(x), \Phi_j(x')\right]=i2\pi \delta_{ij}\delta(x-x').
\end{equation}

The charge mode $\varphi_c$, and the neutral mode  $\varphi_n$ are expressed as follows:
\begin{equation}
\varphi_{c,n}=\frac{1}{\sqrt{2}}\left( \chi_1\pm\chi_2\right),
\end{equation}
which in vector form reads $(\chi_{1},\chi_{2})^T={\hat W}(\varphi_{c},\varphi_{n})^T$,
where the operation $T$ transposes a row vector into a column vector.
The transformation matrix  ${\hat W}$ is given by
\begin{equation}
\mathbf{W}=\frac{1}{\sqrt{2}}\left(\begin{array}{cc}1 & 1\\1 & -1 \end{array}\right),
\label{W}
\end{equation}
and the following relations hold
\begin{eqnarray}
\Phi_1 &=\chi_1 +2\chi_2 &=\frac{1}{\sqrt{2}}\left( 3\varphi_c-\varphi_n \right),
\label{phi1}\\
\Phi_2 &=\chi_2 +2\chi_1&=\frac{1}{\sqrt{2}}\left( 3\varphi_c+\varphi_n \right).
\label{phi2}
\end{eqnarray}
The neutral mode $\varphi_n$ can be interpreted as a difference in the occupation of edge modes corresponding to the first and second $\Lambda$-levels of composite fermions.
The FQHE at $\nu=2/3$ can emerge in two spin polarization states, spin-polarized (p index) with two filled composite fermion $\Lambda$-levels in the same spin state, and spin-uppolarized (u index) with two filled composite fermion $\Lambda$-levels with two opposite spins.
In the unpolarized phase it coincides with the spin density and we will use spin
index $s$ instead of $n$.

\subsection{Description of edges of $\nu=2/3$ state in terms of separate charge and neutral/spin modes}

In a seminal paper by Kane Fisher and Polchinsky [34],  it was suggested that a significant role in the physics of $\nu=2/3 state$  is played by
scatttering between oppositely propagating modes.
However, both polarized and unpolarized phases must be treated on equal footing. This immediately raises a point that the edges in these phases are drastically different: there could be no elastic scattering leading to hybridization of two edge states emerging from the bulk composite fermion picture in the unpolarized phase, where these two edge sttes have opposite spin. We note that true spin system is much more restrictive in this regard compared to and pseudospin layered system, where tunneling is allowed between layers. To resolve this issue, we use observation by Wen \cite{Wen1995} that at strong Coulomb interactions, a spin-charge (neutral-charge) separation occurs. Once we assume this separation takes place, the Luttinger liquid action in both spin-polarized and spin-unpolarized phases has the same form as Kane-Fisher-Polchinsky action.

Therefore we formulate Luttinger liquid action in terms of separate charge and neutral/spin modes. Using matrix $W$ defined Eq. (\ref{W}) the transformed matrix $K$, given by Eq. (\ref{K}) becomes
\begin{equation}
\mathbf{W^TKW}=\left(\begin{array}{cc}3 & 0\\0 & -1 \end{array}\right).
\label{transformedK}
\end{equation}

For the transformed matrix $\hat{V}^{qp}$,
\begin{equation}
V_{cn}=W^T\hat{V}^{qp}W,
\end{equation}
we obtain
\begin{equation}
V_{cn}=\left(\begin{array}{cc}v_c & v_{cn}\\v_{cn} & v_s \end{array}\right),
\label{Vcn}
\end{equation}
where in terms of couplings $V_1$, $V_2$  and $V_3$ in  Eq.(\ref{V}), matrix ements of coupling matrix are given by
\begin{eqnarray}
&v_c=\frac{9}{2}(V_1+V_2+2V_3)\nonumber\\
&v_n=\frac{1}{2}(V_1+V_2-2V_3)\nonumber\\
&v_{cn}=\frac{3}{2}(V_2-V_1)
\end{eqnarray}
For equal diagonal intra-mode couplings $V_1=V_2$ the off-diagonal  terms of the $V_{cn}$ matrix vanish, $v_{cn}=0$,
and the diagonal elements take the form $v_c=9(V_1+V_3)$, $v_n=V_1-V_3$.As was discussed by Wen in Ref. [20],
equal couplings $V_1$ of both composite fermion modes Eq.(\ref{V}) are the consequences of the long-range Coulomb interaction.
[In the absence of the mechanism of electron scattering on impurities this equal coupling allows  separation of charge and neutral modes in polarized phase and charge and spin modes for non-polarized phase.]

In order to distinguish Luttinger liquid modes in \textit{p} and \textit{u} phases, we characterize all modes, including bosonic electron modes, quasiparticle modes, and separated charge and neutral (spin) modes by indices \textit{p} and \textit{u}. The transformed action  for the \textit{p} phase is
\begin{eqnarray}
&S = \frac{1}{4\pi}\int dt \int dx \left[ -3\partial_x \varphi_{pc}(\partial_t+v_c\partial_x) \varphi_{pc} \right.  + \nonumber\\
&\left. \partial_x \varphi_{pn}   (\partial_t- v_n\partial_x) \varphi_{pn} \right],
\label{actionp}
\end{eqnarray}
where $v_c$ and $v_n$ are velocities of charge and neutral modes in polarized phase, correspondingly.
This action coincides with the one
expressed in terms of the charge and neutral fields in [34],  in the case if no electron scattering that leads to the composite fermion (CF) tunneling between the different modes and no coupling between modes takes place. The Luttinger liquid action in \textit{u} phase is similar with spin modes entering instead of neural modes:
\begin{eqnarray}
&S = \frac{1}{4\pi}\int dt \int dx \left[ -3\partial_x \varphi_{uc} \left(\partial_t+ v_c\partial_x \right)\varphi_{uc} \right.  +\nonumber\\
&\left.  \partial_x\varphi_{us}\left(\partial_t- v_s\partial_x \right)\varphi_{us}\right],
\label{S}
\end{eqnarray}
where $v_c$ and $v_s$ are velocities of charge and spin modes, correspondingly. The spin mode determines the spin current in the \textit{u} phase.
In both phases, charge and neutral, or charge and spin modes separate.

The commutation relations for separated charge and neutral modes in the \textit{p} phase are given by
\begin{eqnarray}
&\left[ \partial_x \varphi_{pc}(x), \varphi_{pc}(x')\right]=& i\frac{2\pi}{3}\delta(x-x'),\\
&\left[ \partial_x \varphi_{pn}(x), \varphi_{pn}(x')\right]=&- i2\pi\delta(x-x').
\end{eqnarray}
 In the \textit{u} phase, the commutation relatons for separated charge and spin modes are given by these equations with substitution $pc\rightarrow uc$, $pn\rightarrow us$.
 To acquire non-zero average charge density and current, density of the charge mode $\varphi_{pc}$ is shifted,
$\varphi_{pc}\rightarrow \varphi_{pc}(x,t) +\bar{\varphi}_{pc}$. A non-zero average appears due to a charge current injection,
\begin{equation}
\bar{\varphi}_{pc}=\frac{e\sqrt{2}}{3\hbar}\left(\frac{x}{v_c}-t\right)V,
\label{shift_1}
\end{equation}
where V is the applied voltage. Then the average current carried by the edge is
\begin{equation}
\bar{j}=-\frac{e q_p}{\sqrt{2}\pi}\partial_t\varphi_{pc}= \frac{e^2}{2\pi\hbar}\frac{2}{3}V
\end{equation}
with $q_p=1$ as in Eq.(\ref{K}).
The shift of the charge mode density is described by an addition to the Luttinger liquid action
\begin{equation}
\Delta S = \frac{eV}{4\pi\hbar v_c}\int dt \int dx \sqrt{2} \left(\partial_t+v_c\partial_x\right) \varphi_{pc} ,
\label{shift}
\end{equation}
so that $S(\varphi_{pc}) + \Delta S (\varphi_{pc})= S(\varphi_{pc}- \bar{\varphi}_{pc})$. The case of injection from the unpolarized phase into polarized phase, when $\varphi_{uc}$ is shifted due to applied voltage instead of $\varphi_{pc}$, is described by  Eqs. (\ref{shift_1}-\ref{shift}) with the change $pc\rightarrow uc$.

\subsection{The Luttinger liquid action in the presence of both spin-polarized and spin-unpolarized phases.}

The Luttinger liquid action for the edge states at $\nu=2/3$ consisting of the two phases, polarized \textit{p} and unpolarized \textit{u}, is given by
\begin{eqnarray}
&S = -\frac{1}{4\pi}
\int dt \int dx  \left[ (\partial_x \Phi_u, \partial_x \Phi_p) {\hat{\cal{K}}}^{-1}
\left(\begin{array}{c}\partial_t \Phi_u\\
\partial_t \Phi_p \end{array}\right) +\right. \nonumber\\
&  \left. (\partial_x \Phi_u, \partial_x \Phi_p){\hat{\cal{V}}}\left(\begin{array}{c}\partial_x \Phi_u\\
\partial_x \Phi_p \end{array}\right)\right],
\label{up}
\end{eqnarray}
where 4x4 matrices
 \begin{equation}
{\hat{\cal{K}}}=\left(\begin{array}{cc}
-{\hat{K}} & 0\\
0&
{\hat{K}} \end{array}\right),
\end{equation}
with matrix ${\hat{K}}$ defined by Supplementary Eq.~(\ref{K}) and
\begin{equation}
{\hat{\cal{V}}}=\left(\begin{array}{cc}
{\hat{V}^{e}} & 0\\
0&
{\hat{V}^{e}} \end{array}\right),
\end{equation}
with matrix $\hat{V}^{e}$ is defined by Supplementary  Eq.~(\ref{V}). For convenience, in order to keep the form of relations for the current as described above for both \textit{p} and \textit{u} phases, we reverse the sign of the quasiparticle field $\chi_u\rightarrow -\chi_u$, so that $\Phi_u=K\chi_u$, but $q_u=-1$.

\section{Tunneling current through the quantum point contact between polarized and unpolarized phases}

The point contact $x=0$ tunneling between the polarized and unpolarized phases carried by composite fermions with the same spin polarization is described by the tunneling Hamiltonian ${\cal H}_T$ and current $J_T$:
\begin{eqnarray}
{\cal H}_T= -\tilde{t}\cos{(\Phi_{p1}(0,t)- \Phi_{u1}(0,t))}
\label{tunnelingHS} \\
  J_T=-\tilde{t}\sin{(\Phi_{p1}(0,t)- \Phi_{u1}(0,t))}.
\label{tunnelingcS}
\end{eqnarray}
In order to access the properties of the system, it is instructive to consider the tunneling current [26]  in the strong coupling limit with account for weak backscattering [38].

\subsection{Tunneling current in the strong coupling limit}

Transformation of the two tunneling fields to the right-moving chiral fields is described as follows
\begin{eqnarray}
 &\tilde{\Phi}_{p1}(x,t) =\frac{1}{\sqrt{2}}\left(3\varphi_{pc}(x,t)-\varphi_{pn}(-x,t)\right)\\
&\tilde{\Phi}_{u1}(x,t) =\frac{1}{\sqrt{2}}\left(3\varphi_{uc}(-x,t)-\varphi_{us}(x,t)\right).
\label{rightmoving}
\end{eqnarray}
The two non-tunneling fields orthogonal to fields (\ref{rightmoving}) are described by
\begin{eqnarray}
 &\tilde{\varphi}_{p2}(x,t) =\frac{1}{\sqrt{2}}\left(\varphi_{pc}(x,t)+\varphi_{pn}(-x,t)\right)\\
&\tilde{\varphi}_{u2}(x,t) =\frac{1}{\sqrt{2}}\left(\varphi_{uc}(-x,t)+\varphi_{us}(x,t)\right).
\label{nontunneling}
\end{eqnarray}
The strong coupling boundary conditions of Eq. (\ref{tunnelingH}), $\tilde{t}\rightarrow \infty$ intertwin $\tilde{\Phi}_{p1}(x,t)$ and $\tilde{\Phi}_{u1}(x,t)$ at  $x=0$ crossing:
\begin{equation}
\tilde{\Phi}_{p1}(-0,t)\mp \tilde{\Phi}_{u1}(-0,t) =\mp\left(\tilde{\Phi}_{p1}(+0,t)\mp\tilde{\Phi}_{u1}(+0,t) \right)
\end{equation}
and keep continuous the nontunneling fields $\tilde{\varphi}_{p2}(x,t)$ and  $\tilde{\varphi}_{u2}(x,t)$ and the two dual fields
\begin{eqnarray}
 &\eta_{1}(x,t) =\Phi_{p1}(x,t)\theta(-x)+\Phi_{u1}(x,t)\theta(x)\\
&\eta_{2}(x,t) =\Phi_{u1}(-x,t)\theta(-x)+\Phi_{p1}(x,t)\theta(x),
\label{dual}
\end{eqnarray}
where $\theta(x)$ is the Heaviside step function.
In the presence of voltage V that produces the tunneling current and a density shift via the shift of $\Delta\varphi_{pc}(x,t) \theta(-x)=-\frac{\sqrt{2}}{3}V(t-x/v_{c})$ in Eq.(\ref{shift_1}), we have the following time-dependent averages:
\begin{eqnarray}
&\Delta_t \langle\eta_{1}(x,t)\rangle=-Vt\\
&\Delta_t \langle\eta_{2}(x,t)\rangle=0.
 \end{eqnarray}
Therefore
\begin{eqnarray}
&\Delta_t \langle\tilde{\Phi}_{u1}(x,t)\rangle=\nonumber\\
&\Delta_t \langle\frac{1}{\sqrt{2}}\left(3\varphi_{uc}(-x,t)-\varphi_{us}(x,t)\right)\rangle=-Vt\theta(x)\\
&\Delta_t  \langle\tilde{\varphi}_{u2}(x,t)\rangle= \nonumber\\
&\Delta_t\langle\frac{1}{\sqrt{2}}\left(\varphi_{uc}(-x,t)+\varphi_{us}(x,t)\right)\rangle=0.
 \end{eqnarray}
It then follows
\begin{eqnarray}
&\Delta_t  \langle\varphi_{uc}(-x,t) \rangle=-\frac{\sqrt{2}}{4}Vt\theta(x)\\
&j_{uc}=q_u\sqrt{2}\sigma_0\partial_t \langle \varphi_{uc}(x,t)\rangle=\frac{1}{4\pi}V\theta(-x),
\end{eqnarray}
and
\begin{eqnarray}
&\Delta_t  \langle\varphi_{us}(x,t) \rangle=\frac{\sqrt{2}}{4}Vt\theta(x)\\
&j_{us}(x)=-q_u\sqrt{2}\sigma_0\partial_t \langle \varphi_{us}(x,t)\rangle=\frac{1}{4\pi}V\theta(x).
\end{eqnarray}
Furthermore, the strong coupling boundary conditions yield the following averages  in the presence of voltage for
$\tilde{\Phi}_{p1}(x,t)$ and $\tilde{\varphi}_{p2}(x,t)$:
\begin{eqnarray}
&\Delta_t \langle\tilde{\Phi}_{p1}(x,t)\rangle=\nonumber\\
&\Delta_t \langle\frac{1}{\sqrt{2}}\left(3\varphi_{pc}(x,t)-\varphi_{pn}(-x,t)\right)\rangle=-Vt\theta(-x)\\
&\Delta_t  \langle\tilde{\varphi}_{p2}(x,t)\rangle= \nonumber\\
&\Delta_t\langle\frac{1}{\sqrt{2}}\left(\varphi_{pc}(x,t)+\varphi_{pn}(-x,t)\right)\rangle=-\frac{1}{3}Vt.
 \end{eqnarray}
It then follows for the time-dependent averages and currents in the polarized region:
\begin{eqnarray}
&\Delta_t  \langle\varphi_{pc}(x,t) \rangle=-\frac{\sqrt{2}}{3}Vt\theta(-x)-\frac{\sqrt{2}}{12}Vt\theta(x)\\
&j_{pc}=-q_p\sqrt{2}\sigma_0\partial_t \langle \varphi_{pc}(x,t)\rangle=\nonumber\\
&\frac{1}{2\pi}V\left(\frac{2}{3}\theta(-x)+\frac{1}{6}\theta(x)\right),
\end{eqnarray}
and
\begin{eqnarray}
&\Delta_t  \langle\varphi_{pn}(-x,t) \rangle=\frac{\sqrt{2}}{4}Vt\theta(x)\\
&j_{pn}(x)=-q_p\sqrt{2}\sigma_0\partial_t \langle \varphi_{pn}(x,t)\rangle=\nonumber\\
&-\frac{1}{4\pi}V\theta(-x).
\end{eqnarray}
We observe that tunneling charge current $\langle J_{cT}\rangle$ from polarized region coincides with tunneling spin current current:
\begin{eqnarray}
&\langle J_{cT}\rangle=j_{pc}(x<0)-j_{pc}(x>0)=\frac{V}{4\pi}=\nonumber\\
&=\langle J_{sT}\rangle= j_{us}(x>0)=\langle J_{T}\rangle
\end{eqnarray}
and is opposite in sign to reflected neutral current in polarized region.

\subsection{Reduction of tunneling current due to backscattering}

Quasiparticle tunneling results in backscattering that reduces the tunneling current. Backscattering is associated with a sudden change at $t=t_0$ of the strong coupling boundary conditions
\begin{equation}
\frac{1}{2}\sum_{\pm}\left(  \tilde{\Phi}_{p1}(\pm0,t_0) -\tilde{\Phi}_{u1}(\pm0,t_0)\right)=2\pi n,
\end{equation}
from integer $n=0$ to integer $n=1$ that both minimize $-\tilde{t}\cos{(\Phi_{p1}(0,t)- \Phi_{u1}(0,t))}$. This leads to a jump in the fields
\begin{equation}
\left(\eta_1(x,t_0)- \eta_2(x,t_0)\right)\vert_{x=-0}^{+0}=-4\pi.
\end{equation}
Since
\begin{equation}
\left(\eta_1(x,t_0)+\eta_2(x,t_0)\right)\vert_{x=-0}^{+0}=0,
\end{equation}
the corresponding jumps in the fields are given by
\begin{equation}
\Delta\eta_2(+0,t_0)=2\pi= -\Delta\eta_1(+0,t_0)
\end{equation}
that lead to the following changes
\begin{equation}
\Delta\left(3\varphi_{uc}(-x,t_0)-\varphi_{us}(x,t_0)\right)=-\sqrt{2}2\pi\theta(x)
\end{equation}
At the same time,
\begin{equation}
\Delta\left(\varphi_{uc}(-x,t_0)+\varphi_{us}(x,t_0)\right)=0.
\end{equation}
Then
\begin{equation}
\Delta\varphi_{uc}(-x,t_0)=-\frac{\pi}{\sqrt{2}}\theta(x)
\end{equation}
and the change of the charge density $\delta\rho_{uc}(x,t_0)=-\frac{1}{2}\delta(x)$.
Analogously, the change in spin mode
 \begin{equation}
\Delta\varphi_{us}(x,t_0)=\frac{\pi}{\sqrt{2}}\theta(x),
\end{equation}
and the change of the spin density
 \begin{equation}
\delta\rho_{us}(x,t_0)=\frac{q_u}{\sqrt{2}\pi}\partial_x\Delta\varphi_{us}=-\frac{1}{2}\delta(x).
\end{equation}
Furthermore, from $\Delta\eta_2(+0,t_0)=2\pi$ we get the change
\begin{equation}
\Delta\left(3\varphi_{pc}(x,t_0)-\varphi_{pn}(-x,t_0)\right)= 2\sqrt{2}\pi\theta(x).
\end{equation}
Using the continuity condition
\begin{equation}
\Delta\left(\varphi_{pc}(x,t_0)+\varphi_{pn}(-x,t_0)\right)=0
\end{equation}
we find
\begin{equation}
\Delta\varphi_{pc}(x,t_0)=\frac{\pi}{\sqrt{2}}\theta(x);\hspace{4mm}
\Delta\varphi_{pn}(-x,t_0)=-\frac{\pi}{\sqrt{2}}\theta(x)
\end{equation}
and for the change in the charge density and in the density of the neutral mode
\begin{eqnarray}
&\delta\rho_{pc}=\frac{q_p}{\sqrt{2}\pi}\partial \Delta\varphi_{pc}=\frac{1}{2}\delta(x)\\
&\delta\rho_{pn}=\frac{1}{2}\delta(x)
\end{eqnarray}
As was shown in [62], the calculated transferred charge enables us to identify the charge of backscattering quasiparticles q=e/2. Surprisingly, this charge is different from q=e/3 which one expects from [34] . This difference stems from the absence of tunneling between two edge states of different spin polarizations in the polarized and unpolarized regions.

These calculations demonstrate that for tunneling through the QPC, in which one CF edge on one side of the QPC is scattered into one CF edge on the other side of the QPC the quasiparticle charge is $e/2$ and the conductance saturates at $G=e^2/2h$. This occurs in tunneling between both spin-like states of the polarized and unpolarized region or when the confining potential is engineered to allow only one mode to tunnel between two polarized or two unpolarized regions.

\end{document}